# POLYPHONIC SOUND EVENT DETECTION BASED ON CONVOLUTIONAL RECURRENT NEURAL NETWORKS WITH SEMI-SUPERVISED LOSS FUNCTION FOR DCASE CHALLENGE 2020 TASK 4

Technical Report

*Nam Kyun* Kim[1] *and Hong Kook Kim*[1,2]

[1]School of Electrical Engineering and Computer Science, [2]AI Graduate School
Gwangju Institute of Science and Technology
123 Cheomdangwagi-ro, Gwangju 61005, Korea
{skarbs001, hongkook}@gist.ac.kr

## ABSTRACT

This report proposes a polyphonic sound event detection (SED) method for the DCASE 2020 Challenge Task 4. The proposed SED method is based on semi-supervised learning to deal with the different combination of training datasets such as weakly labeled dataset, unlabeled dataset, and strongly labeled synthetic dataset. Especially, the target label of each audio clip from weakly labeled or unlabeled dataset is first predicted by using the mean teacher model that is the DCASE 2020 baseline. The data with predicted labels are used for training the proposed SED model, which consists of CNNs with skip connections and self-attention mechanism, followed by RNNs. In order to compensate for the erroneous prediction of weakly labeled and unlabeled data, a semi-supervised loss function is employed for the proposed SED model. In this work, several versions of the proposed SED model are implemented and evaluated on the validation set according to the different parameter setting for the semi-supervised loss function, and then an ensemble model that combines five-fold validation models is finally selected as our final model.

*Index Terms*— Polyphonic sound event detection, semi-supervised loss function, self-attention mechanism

## 1. INTRODUCTION

Sound event detection (SED) is a task of classifying individual sound events in diverse sound environments and detecting the onset and offset times of each detected sound event. SED can support understanding of multimedia contents in more detail and monitor acoustic scenes for various applications such as audio surveillance [1], audio monitoring in smart cities [2], life assistance and health care [3], and so on.

While deep learning-based SED systems that are trained using a large amount of strong labeled data provide good performance, reliably annotating temporal extents of event occurrence is very difficult and time-consuming due to human errors [4]. Since 2017, the Detection and Classification of Acoustic Scenes and Events (DCASE) Challenge has featured an SED task to deal with weakly labeled data that do not have any time stamps for given audio events [5, 6].

The DCASE Challenge 2020 Task 4, which is the follow-up to the DCASE Challenge 2019 Task 4, focuses on detecting polyphonic sound events, where the SED model should be developed using three different datasets: a weakly labeled dataset, unlabeled in-domain dataset, and a few strongly labeled synthetic data. According to the results of the DCASE Challenge 2019 Task 4, some of top-ranked models were based on a mean teacher model [4] trained by both weakly labeled data and unlabeled data with consistency regularization. Specifically, the mean teacher model aims to learn the two same structures, while the teacher model aims to help the student model during training and its model parameters are updated by the exponential moving average of the student model parameters.

Instead of modifying the mean teacher model, we propose an SED model that uses a semi-supervised loss function [7] that combines a supervised loss function for strongly labeled data and an unsupervised loss function for weakly-labeled or unlabeled data. For this loss function, we need to give a label and time stamp for each weakly labeled or unlabeled dataset. This is performed by using the mean teacher model prior to training the proposed SED model. The proposed SED model is composed of convolutional neural networks (CNNs) with skip connection, followed by convolutional block attention, which carries out a self-attention mechanism. Finally, recurrent neural networks (RNNs) are connected to the prediction layer for detecting events and time-stamps.

## 2. DATASET

The dataset of DCASE 2020 Challenge Task 4 consists of three types in the training process: 1) weakly labeled training dataset (without timestamp), 2) unlabeled in-domain training dataset without any label, and 3) strongly labeled synthetic dataset. The weakly labeled and the unlabeled in-domain training datasets are taken from the AudioSet [8], but the strongly labeled synthetic dataset is generated using the Scaper soundscape synthesis and augmentation library [9]. The weakly labeled training dataset contains 1,578 audio clips with weak annotation only, where there are 2,244 class occurrences. The unlabeled in-domain training dataset and the strongly labeled dataset contain 14,412 and 2,584 audio clips, respectively. Each audio clip is stored as both mono and stereo-channel signals that are sampled at 44.1 kHz with a maximum duration of 10 seconds.

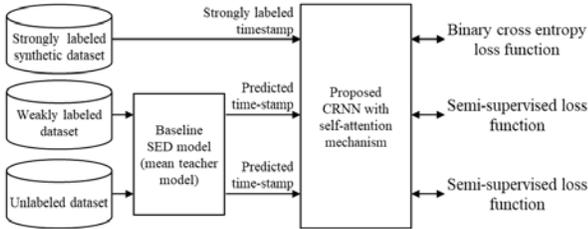

Figure 1: Procedure of the proposed semi-supervised SED method.

For a given dataset, we first take mono-channel signals and resample them from 44.1 to 16 kHz. After that, each resampled audio signal is segmented into consecutive frames of 512 samples with 255 samples of hop length. Then, a 512-point fast Fourier transform (FFT) is applied to each separated signal, and a 128-dimensional mel-filterbank analysis is performed for each frame. Since each 10-second audio clip is represented by 628 frames, the dimension of input feature for the SED model is 1×628×128. Note here that zero padding is applied to the audio clips that are shorter than 10 seconds. Finally, the extracted mel-spectrogram features are normalized by the global mean and the standard deviation over all the training audio clips.

## 3. PROPOSED SED METHOD

Fig. 1 shows a block diagram of the proposed SED method. As shown in the figure, we first train a mean teacher model that is identical to the baseline of the DCASE Challenge 2020 Task 4 [4]. Next, time stamps for each training audio clip are decoded using the trained mean teacher model. In other words, each of the weakly-labeled or unlabeled audio clips is inferred using the mean teacher model, resulting in a (157×10)-dimensional time-stamp map that will be used for the target of the SED model. For the strongly labeled dataset, we just use the given label and time-stamp for the target. After that, such target time-stamps for all the training audio clips are brought to train the proposed SED model.
　　Fig. 2 shows the network architecture of the proposed SED model that is composed of a series of convolutional layer blocks with skip connections and three convolutional attention blocks, followed by RNNs and prediction layer. To train the proposed model, the semi-supervised loss function is employed to accommodate the ambiguity of the time-stamps predicted for the weakly-labeled and unlabeled datasets, as described in Fig. 1.

### 3.1. Proposed CRNN architecture

This subsection explains the network architecture of the proposed SED model in detail. As shown in Fig. 2, the input feature whose dimension is 1×628×128 is applied to a stem block that consists of two convolutional blocks. Each convolutional block contains two CNN layers and a 2×2 average pooling layer. Following the stem block, the skip connected convolution (SCC) block is comprised of five convolutional blocks. The upper part of the SCC block is composed of three convolutional blocks, each of which contains two CNN layers. In addition, the lower part of the SCC block is composed of two convolutional blocks that contain two CNN layers each with a (1×2) average pooling layer. The outputs of the first and second convolutional blocks of the upper part of the SCC block are concatenated into the second and first convolutional blocks of the lower part of the SCC block, as shown in the figure. Moreover, the convolutional block attention module (CBAM) [10] is applied to the three convolutional blocks of the lower part of the SCC block. After that, the feature map extracted from the CNN block is average pooled on the frequency axis and connected to the recurrent block. The recurrent block is composed of two bi-directional long short-term memory (BLSTM) networks with 128 hidden nodes each. Finally, a fully connected (FC) layer is used as the output layer, whose dimension is 157×10. A rectified linear unit (ReLu) is used as the activation function for all the layers except the output layer, while the sigmoid function is used for the output layer.

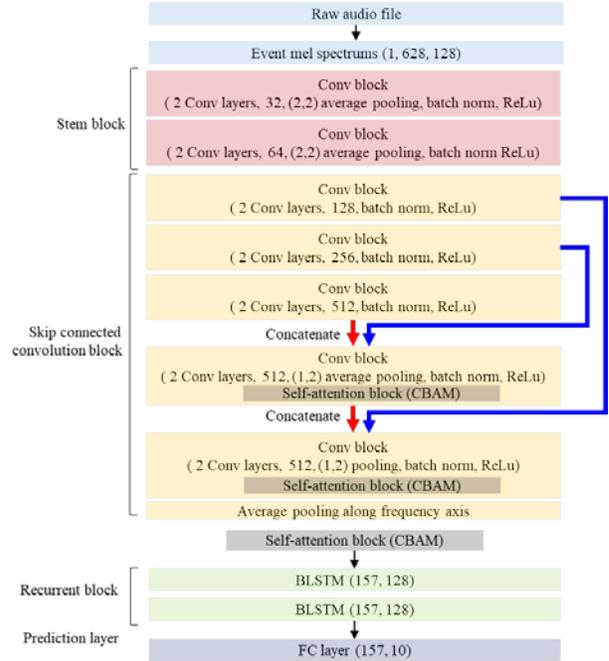

Figure 2: Network architecture of the proposed SED model.

### 3.2. Data augmentation

It is known that data augmentation can improve prediction accuracy when the amount of training data is limited as in this DCASE Challenge. In this work, we perform both the mix-up data augmentation [11] and the spec augmentation [12]. Mix-up augmentation creates a new training sample by mixing a pair of two randomly chosen training samples. Spec augmentation is an effective approach which shows significant performance improvement in acoustic speech recognition recently. It replaces values by zeros in randomly chosen time-frequency bands. Then, a global noise level is computed to add a random Gaussian noise with zero mean and unit standard deviation matching this global noise level.

### 3.3. Semi-supervised loss function

As mentioned earlier, the proposed model is trained by using three different datasets: strongly labeled (S), weakly labeled (W), and unlabeled (U) datasets. Among them, each of the weakly-labeled and unlabeled audio clips is labeled using the mean teacher model.

Detection and Classification of Acoustic Scenes and Events 2020                                                                                                ChallengeTable 1: Comparison of performance metrics on the validation set.

| Method | | Event-based metrics | | PSDS | PSDS cross-trigger | PSDS macro |
|---|---|---|---|---|---|---|
| | | F1-score (%) | Error rate | | | |
| Single model | Baseline | 37.24 | 1.39 | 0.610 | 0.524 | 0.442 |
| | Proposed model ($\beta_s$ =0.3) | 41.92 | 1.17 | 0.698 | 0.632 | 0.537 |
| | Proposed model ($\beta_s$ =0.5) | 42.46 | 1.14 | 0.701 | 0.637 | 0.544 |
| | Proposed model ($\beta_s$ =0.7) | 41.18 | 1.15 | 0.700 | 0.636 | 0.541 |
| | Proposed model ($\beta_s$ =0.9) | 42.41 | 1.13 | 0.697 | 0.632 | 0.532 |
| 5-fold ensemble model | Proposed model ($\beta_s$ =0.3) | 44.60 | 1.05 | 0.724 | 0.666 | 0.560 |
| | Proposed model ($\beta_s$ =0.5) | 45.23 | 1.05 | 0.725 | 0.667 | 0.562 |
| | Proposed model ($\beta_s$ =0.7) | 44.53 | 1.04 | 0.725 | 0.667 | 0.562 |
| | Proposed model ($\beta_s$ =0.9) | 43.96 | 1.07 | 0.721 | 0.665 | 0.558 |

However, these predicted labels could be erroneous because the performance of the mean teacher model is not perfect, and the unreliability of predicted labels could affect the training of the proposed SED model.

To remedy this problem, a semi-supervised loss function is defined with the help of the loss function in [7], such as

$$L_{semi} = -\frac{1}{M} \sum_{s \in \{S,W,U\}} \sum_{k=1,c=1}^{K,C} \{\bar{y}_s^{k,c} \log \hat{y}_s^{k,c} + (1 - \bar{y}_s^{k,c}) \log(1 - \hat{y}_s^{k,c})\} \quad (1)$$

$$\bar{y}_s^{k,c} = \beta_s y_s^{k,c} + (1 - \beta_s) \hat{y}_s^{k,c} \quad (2)$$

where $M$ is the total number of data in training set, $\hat{y}_s^{k,c}$ is $k$-time frame and $c$-th class output value when an audio clip belonging to the dataset $s$ is inputted, and $y_s^{k,c}$ is the target value corresponding to $\hat{y}_s^{k,c}$. Thus, $\beta_s$ is a parameter to control the influence of weakly labeled and unlabeled data; thus, $\beta_s = 1$ for $s \in S$, which results in a binary cross-entropy loss function, and $0 < \beta_s < 1$ for $s \in \{W, U\}$. In Eq. (1), $C$ and $K$ are the numbers of classes and time dimensions. In this work, $C$=10 and $K$=157.

## 4. EXPERIMENT RESULTS

The proposed SED models with different $\beta_s's$ were learnt by using training datasets described in Section 2. For training the baseline and proposed SED models, each batch was designed to include a combination of unlabeled, weakly labeled, and strongly labeled synthetic audio clips, with 1/4, 1/2, and 1/4 portions, respectively. In particular, each of the proposed SED models was trained by 5-fold cross validation by dividing all the data in the training set into 5 folds, where 4 out 5 folds were used for training and the remaining fold was used for validation. Finally, an ensemble classifier was obtained by linearly combining 5-fold models for a given $\beta_s$. Note here that the baseline was trained according to the recipe given by the challenge. However, the Adam optimizer was used for training the proposed models, where the early stopping technique was employed and the dropout was applied with a rate of 0.3. The learning rate was set to 0.0009, and it was reduced by the simple learning rate schedule by a constant factor when the performance metric using mean square error (MSE) plateaus on the validation set (commonly known as ReduceLRonPlateau in Pytorch).

Table 1 compares the performance of the validation set predicted by the baseline and proposed models with different parameter settings, where the performance was measured by the event-based metric (macro-average F1-score and error rate (ER)) and the polyphonic sound detection score (PSDS) [13]. The validation set was composed of 1,168 audio clips with strong labels including time-stamps. We first obtained four different proposed models by setting $\beta_s$ from 0.3 to 0.9 at a step of 0.2. The results at the second to fifth row of the table were obtained by averaging all the 5-fold validation results. Compared to the baseline, all the models with different $\beta_s's$ were better than the baseline in all the performance metrics. Especially, the proposed model with $\beta_s = 0.5$ achieved higher F1-score and lower ER by 5.52 and 0.25, respectively, than the baseline.

Next, we combined 5-fold models into a single classifier. In this case, the ensemble model with $\beta_s$=0.9 was the best in terms of event-based metrics, while the ensemble model with $\beta_s$=0.5 provided the highest PSDS scores for all the ensemble models. Compared to the baseline, the ensemble model increased F1-score by 7.99 and decreased the ER by 0.35. Also, the PSDS score was increased by 0.115 when the ensemble classifier of the proposed 5-fold models with $\beta_s$=0.5.

## 5. CONCLUSION

This report proposed a semi-supervised learning method for polyphonic SED when the training data were strongly labeled, weakly labeled, or unlabeled. To this end, weakly labeled or unlabeled data were each first predicted by using the mean teacher model. The data with predicted labels were used for training the proposed SED model that consisted of CNNs with skip connections and self-attention mechanism, followed by RNNs. In order to compensate for the erroneous prediction of weakly labeled and unlabeled data, a semi-supervised loss function was employed for the proposed SED model.

From the performance evaluation of the proposed SED model on the validation set, it was shown that the ensemble SED model after 5-fold validations achieved an F1-score 7.99 higher and an error rate 0.35 lower than the baseline of the DCASE Challenge 2020 Task 4. Also, the PSDS improvement of 0.115 was achieved by the ensemble classifier, compared to the baseline.

## 6. ACKNOWLEDGMENT

This work was supported in part by an Institute of Information & communications Technology Planning & Evaluation (IITP) grant funded by the Korea government (MSIT) (No. 2019-0-01767, Development of Machine learning-Based Acoustic Intelligence Technology for Disaster Response Using Drones) and (No. 2019-0-01842, Artificial Intelligence Graduate School (GIST)).

Detection and Classification of Acoustic Scenes and Events 2020     Challenge## 7. REFERENCES

[1] G. Valenzise, L. Gerosa, M. Tagliasacchi, F. Antonacci, and A. Sarti, "Scream and gunshot detection and localization for audio-surveillance systems," in *Proc. AVSS*, London, UK, Sept. 2007, pp. 21–26.

[2] J. P. Bello, C. Silva, O. Nov, R. L. DuBois, A. Arora, J. Salamon, C. Mydlarz, and H. Doraiswamy, "SONYC: a system for the monitoring, analysis and mitigation of urban noise pollution," *Communications of the ACM*, vol. 62, no. 2, pp. 68–77, Feb. 2019.

[3] Y. Zigel, D. Litvak, and I. Gannot, "A method for automatic fall detection of elderly people using floor vibrations and soundproof of concept on human mimicking doll falls," *IEEE Transactions on Biomedical Engineering*, vol. 56, no. 12, pp. 2858–2867, Aug. 2009.

[4] N. Turpault, R. Serizel, A. Parag Shah, and J. Salamon, "Sound event detection in domestic environments with weakly labeled data and soundscape synthesis," in *Proc. DCASE Workshop*, New York, NY, Oct. 2019, pp. 253–257.

[5] R. Serizel, N. Turpault, H. Eghbal-Zadeh, and A. Parag Shah, "Large-scale weakly labeled semi-supervised sound event detection in domestic environments," in *Proc. DCASE Workshop*, Surrey, UK, Nov. 2018, pp. 19–23.

[6] A. Mesaros, T. Heittola, A. Diment, B. Elizalde, A. Shah, E. Vincent, B. Raj, and T. Virtanen, "DCASE 2017 Challenge setup: tasks, datasets and baseline system," in *Proc. DCASE Workshop*, Munich, Germany, Nov. 2017, pp. 85–92.

[7] E. Fonseca, M. Plakal, D. P. Ellis, F. Font, X. Favory, and X. Serra, "Learning sound event classifiers from web audio with noisy labels," in *Proc. ICASSP*, Brighton, UK, May 2019, pp. 21–25.

[8] J. F. Gemmeke, D. P. W. Ellis, D. Freedman, A. Jansen, W. Lawrence, R. C. Moore, M. Plakal, and M. Ritter, "Audio set: an ontology and human-labeled dataset for audio events," in *Proc. ICASSP*, New Orleans, LA, Mar. 2017, pp. 776–780.

[9] J. Salamon, D. MacConnell, M. Cartwright, P. Li, and J. P. Bello, "Scaper: a library for soundscape synthesis and augmentation," in *Proc. WASPAA*. IEEE, New Paltz, NY, Oct. 2017, pp. 344–348.

[10] S. Woo, J. Park, J.-Y. Lee, and I. S. Kweon, "CBAM: convolutional block attention module," in *Proc. ECCV*, Munich, Germany, Sept. 2018, pp. 3–19.

[11] H. Zhang, M. Cisse, Y. N. Dauphin, and D. Lopez-Paz, "Mixup: beyond empirical risk minimization," arXiv preprint arXiv:1710.09412, 2017.

[12] D. S. Park, W. Chan, Y. Zhang, C.-C. Chiu, B. Zoph, E. D. Cubuk, and Q. V. Le, "Specaugment: a simple data augmentation method for automatic speech recognition," arXiv preprint arXiv:1904.08779, 2019.

[13] Ç. Bilen, G. Ferroni, F. Tuveri, J. Azcarreta, and S. Krstulović, "A framework for the robust evaluation of sound event detection," in *Proc. ICASSP*, Barcelona, Spain, May 2020, pp. 61–65.